\documentclass[
aps,
prl,
reprint,
superscriptaddress,
amsmath,
amssymb,
]{revtex4-1}

\usepackage{color}
\usepackage{amsmath}
\usepackage{amssymb}
\usepackage{amsfonts}
\usepackage{amsthm}
\usepackage{float}
\usepackage{bm}
\usepackage{latexsym}
\usepackage{hyperref}
\usepackage{multirow}
\usepackage[capitalize]{cleveref}
\usepackage{graphicx}
\hypersetup{
    colorlinks,
    citecolor=blue,    filecolor=blue,
    linkcolor=blue,     urlcolor=blue
}

\begin{document}
\title{Staged laser acceleration of high quality protons from a tailored plasma}

%Self-truncated monoenergetic proton acceleration on a tailored density plasma}
\author{Y. Wan}
\affiliation{Department of Engineering Physics, Tsinghua University, Beijing 100084, China}
\affiliation{Department of Physics of Complex Systems, Weizmann Institute of Science, Rehovot 7610001, Israel}

\author{I. A. Andriyash}
\affiliation{Department of Physics of Complex Systems, Weizmann Institute of Science, Rehovot 7610001, Israel}

\author{J. F. Hua}
\affiliation{Department of Engineering Physics, Tsinghua University, Beijing 100084, China}

\author{C. -H. Pai}
\affiliation{Department of Engineering Physics, Tsinghua University, Beijing 100084, China}

\author{W. Lu}
\email[]{weilu@tsinghua.edu.cn}
\affiliation{Department of Engineering Physics, Tsinghua University, Beijing 100084, China}

\author{W. B. Mori}
\affiliation{University of California Los Angeles, Los Angeles, CA 90095, USA}

\author{C. Joshi}
\affiliation{University of California Los Angeles, Los Angeles, CA 90095, USA}

\author{V. Malka}
\affiliation{Department of Physics of Complex Systems, Weizmann Institute of Science, Rehovot 7610001, Israel}
\affiliation{Laboratoire d'Optique Appliqu\'{e}e, ENSTA-CNRS-Ecole Polytechnique, UMR7639, 91761 Palaiseau, France}

\date{\today}

\begin{abstract}
A new scheme of proton acceleration from a laser-driven near-critical-density plasma is proposed. Plasma with a tailored density profile allows a two-stage acceleration of protons. The protons are pre-accelerated in the laser-driven wakefields, and are then further accelerated by the collisionless shock, launched from the rear side of the plasma. The shock has a small transverse size, and it generates a strong space-charge field, which defocuses protons in such a way, that only those protons with the highest energies and low energy spread remains collimated. Theoretical and numerical analysis demonstrates production of high-energy proton beams with few tens of percents energy spread, few degrees divergence and charge of few nC. This  scheme indicates the efficient generation of quasi-monoenergetic proton beams with energies up to several hundreds of MeV with PW-class ultrashort lasers.
\end{abstract}
\maketitle

Over the past decades, many new exciting applications of the interactions of ultraintense lasers with matter have been developed \cite{mourou2006optics}. One very attractive process is the laser-driven ion acceleration \cite{daido2012,macchi2013}, which has great potential for applications of compact ion sources to ultrafast radiography \cite{borghesi2002PIprl, li2006pi}, radiotherapy \cite{bulanov2002CTpra,malka2004ct}, etc. Many applications require ion beams with narrow energy spread, low divergence and sufficient charges. Presently, various mechanisms of ion acceleration from laser plasmas are being explored including target-normal sheath acceleration \cite{snavely2000TNSA}, radiation pressure acceleration \cite{esirkepov2004, macchi2005, zhang2007a, robinson2008, klimo2008, yan2008, wan2016}, breakout afterburner \cite{albright2007boa,yin2007}, laser driven shock acceleration \cite{silva2004, haberberger2012, fiuza2012prl} and magnetic vortex acceleration \cite{nakamura2010,helle2016} etc. These mechanisms in turn employ a variety of targets. Among these, moderate or near-critical density (NCD) plasmas have received a great deal of attention \cite{kuznetsov2001, bulanov2005, willingale2006, yogo2008, fukuda2009, bulanov2010pop, nakamura2010, helle2016, sharma2018high}, particularly for exciting shock-like electrostatic structures capable of accelerating ions to high energies. In this case, the resulting beams may contain large ion numbers ($\sim10^9$), but typically have rather broad spectra, and lack angular collimation. Furthermore, the generation of high-charge low-divergence and narrow-energy spread ion beams of hundred plus MeV with currently available PW-class ultrashort lasers remains a critical challenge to date.

In this Letter, we propose and discuss a new concept to produce a narrow-divergence quasi-monoenergetic beam of protons with hundreds of MeV energies and few nC charges. This concept employs an ultrashort laser with power from a few hundreds TW to a PW, focused into a sharply tailored NCD target. The key feature that allows all these challenges to be simultaneously met is the plasma design that enables the excitation of a wakefield and a shock with a small transverse size. Protons are pre-accelerated by the laser-driven wakefield and then further accelerated by the collisionless shock. The narrow energy spectrum and small angular divergence results from a strong gate effect produced by the co-moving defocusing electrostatic field of the shock itself. In this two-stage proton acceleration process, laser first travels through the uniform density region of the NCD plasma where it drives a strongly nonlinear wake. The longitudinal field of the wake traps hot plasma electrons. These electrons are radially compressed by the azimuthal magnetic field. Upon reaching the sharp density downramp, they generate a transversely narrow shock where further acceleration occurs. The resulting proton beam contains a high charge, and has excellent spectral and angular quality.

To develop this concept, we perform a detailed study based on three-dimensional numerical simulations and support it with a theoretical model. For convenience and model scalability, we adopt dimensionless units, normalizing time and length to $1/\omega_0$ and $c/\omega_0$ respectively, where $\omega_0$ and $c$ are laser frequency and light speed in vacuum respectively. Plasma density is normalized to $n_c=m_e\omega_0^2/4\pi e^2$, and the laser field is described by the normalized amplitude $a_0=eE_0/m_ec\omega_0$, where $m_e$, $e$ are electrons rest mass and charge respectively. We set the initial time, $t_0=0$, to the moment when laser reaches the target front side. To model laser plasma interactions, we use the 3D relativistic Particle-In-Cell (PIC) code OSIRIS \cite{fonseca2002}.

\begin{figure}[!ht]
\centering
  \includegraphics[width=\linewidth]{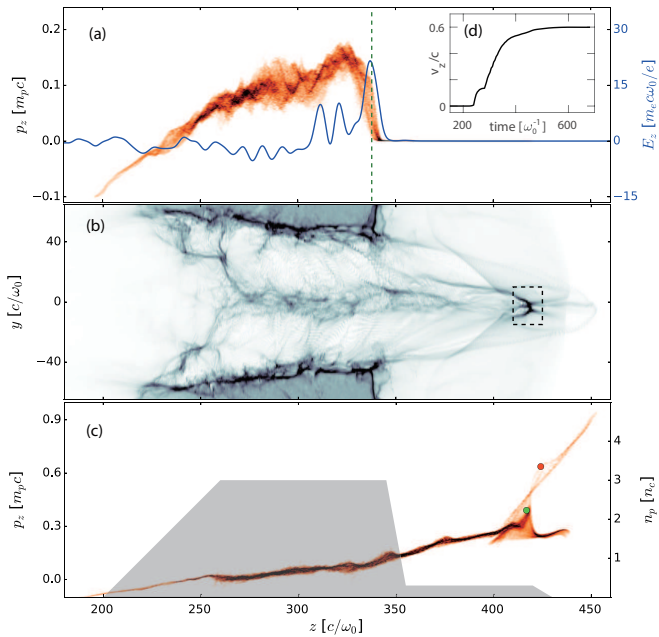}
 \caption{\label{3Dexample_stages} (a) The averaged longitudinal electric field $E_z$ and proton ($z-P_z$) phase space within the region $|x,\ y|$ $\leq 10\, c/\omega_0$ at $t = 220\, \omega_0^{-1}$, where the laser peak position is marked with the dashed green line.
 (b) The $z-y$ slice (at $x=0$) distribution of proton density $n_\text{prot}$ at $t = 440\, \omega_0^{-1}$.
 (c) The ($z-P_z$) phase space of protons picked within the region $|x,\ y|$ $\leq 10\ c/\omega_0$, at the same time as (b). The initial plasma density profile is plotted using gray filled curve.
 (d) The time evolution of a tracked proton's longitudinal velocity.}
\end{figure}

In our simulations, the electron-proton plasma is initially cold and fully ionized, and its density profile is shown by the gray filled curve in Fig.~\ref{3Dexample_stages}(c). The plasma starts at 200~$c/\omega_0$ with a linear ramp  $60\, c/\omega_0$, reaching its maximum value of $n_{1} = 3 n_c$. The entrance ramp is followed by the plateau $L_1 = 85\,c/\omega_0$, that ends with a sharp linear downramp of $d = 10\, c/\omega_0$, where density falls to $n_{2}=0.3n_c$, then remains constant along the distance $L_2 = 65\,c/\omega_0$, and finally ends with another downramp, $10\, c/\omega_0$. We consider a circularly polarized laser focused at plasma surface at $z=200\, c/\omega_0$ with $a_0=42$, spot diameter (FWHM) $30\, c/\omega_0$, and pulse duration (FWHM) $70\, \omega_0^{-1}$.  To ensure validity of the numerical model, we use the numerical grid of the size of $N_x=N_y=320$, $N_z=4000$, which accurately resolves the space as, $\Delta y=\Delta x= 0.5\;c/\omega_0$ and $\Delta z=0.25 c/\omega_0$. In simulations, plasma is initiated with 16 macroparticles per species per cell. For a reference case of a 30 fs, 800nm laser, the corresponding length of the plasma is about 28 $\mu m$, and laser power is close to 1 PW.

As mentioned earlier, in the first interaction stage, laser propagates through the NCD plasma because of the relativistic reduction of the plasma frequency, and drives a strongly nonlinear wake. The accelerating electric component of the wakefield for protons is the strongest within the first half wake period, while the following half (the decelerating phase) of the wake is suppressed by beam loading effect\cite{katsouleas1987, tzoufras2008}, i.e. is screened by the plasma electrons self-injected into the wake (see Fig.~\ref{3Dexample_stages}(a)). For protons this accelerating field is strong enough to impart a significant longitudinal momentum (see $z-P_z$ phase space in Fig.~\ref{3Dexample_stages}(a)).
Inside the laser-generated channel, induced magnetic field confines these thermal trapped plasma electrons transversely near the channel axis \cite{bulanov2010pop, debayle2017}. The effect of electron heating and collimation is enhanced in the case of circular laser polarization \cite{liu2013}. When passing the downramp, these electrons produce a strong pressure, which, together with the magnetic pressure, launch a strong shock with roughly the same small radius as the hot electrons flux \cite{nakamura2010,fiuza2013pop}. In Fig.~\ref{3Dexample_stages}(b) we show the spatial structure of the shock at the later time, when it is already formed. The density in the shock can reach $\sim 30\,n_c$, and its transverse size is extremely small, $\sim 8\, c/\omega_0$ (black dashed box). The longitudinal proton phase distribution in Fig.~\ref{3Dexample_stages}(c), demonstrates that protons reflected by the shock can reach $p_z \simeq 0.8\, m_p c$. Physically this second stage of energy gain is due to shock
acceleration \cite{silva2004,haberberger2012,fiuza2012prl}. As a result of such cascaded process, protons obtain high energies with a large positive chirp.

The two acceleration stages, can be also distinguished in the time evolution of the proton velocity shown in Fig.~\ref{3Dexample_stages}(d). First the protons get accelerated in the wakefield potential during $t_\text{lwfa}\sim 40\, \omega_0^{-1}$, and reaches a high velocity, $\simeq 0.15\,c$. Second, the shock propagates quasi-inertially, and the protons at its front are boosted up to velocities $\simeq  0.6 c$, which corresponds to 240 MeV. Note, that as mentioned before, the circular polarization of laser improves the hot electrons capturing, thus improving the overall acceleration performance \cite{liu2013}, so that in the linearly polarized laser case, proton energy only reaches 150 MeV.

\begin{figure*}[!ht]
\centering
 \includegraphics[width=\linewidth]{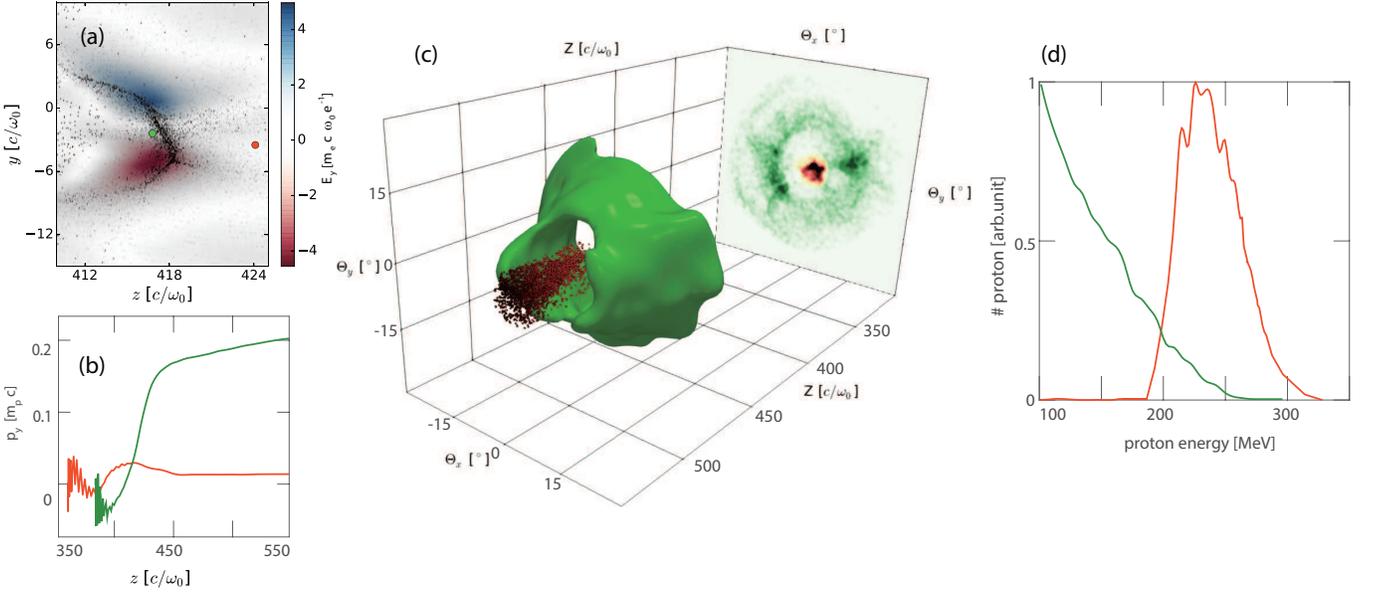}
 \caption{\label{3D_defoc} (a) The $z-y$ slice (at $x=0$) distribution of transverse electric field $E_y$ and ($z-y$) proton phase space at $t = 440\ \omega_0^{-1}$.
 (b) The ($z-P_y$) trajectories of two sample protons traced from the positions given by the colored dots in (a) and Fig.~\ref{3Dexample_stages}(b).
 (c) The 3D trace space ($z-\theta_x-\theta_y$) and its 2D projection ($\theta_x-\theta_y$) for protons with the energies $\epsilon_\text{prot}\geq 80$ MeV at $t = 600\ \omega_0^{-1}$.
 (d) The energy spectra of protons selected in (c).
 In (c)(d), the orange and green correspond to protons with divergence angles less than 100 mrad and more than 100 mrad, respectively.}
\end{figure*}

Besides the longitudinal acceleration, the key feature of the proposed scheme is the strong transverse electrostatic field of the radially compressed shock (see Fig.~\ref{3D_defoc}(a)). The protons within or close to the shock experience strong defocusing, and acquire a significant transverse divergence, however the faster particles located in the front of the shock, do not feel this field, and remain well collimated. This effect is demonstrated in Fig.~\ref{3D_defoc}(b) by the ($z-P_y$) trajectories of two sample protons traced as colored dots in Fig.~\ref{3Dexample_stages}(c) and Fig.~\ref{3D_defoc}(a). One can clearly see that the proton starting ahead of the shock with higher energy (orange line) tends to move along the axis without being deflected, while the particle located closer to the shock (green line) is expelled in the transverse direction, gaining a large transverse momentum $0.2m_pc$.

The above described transverse defocusing process allows the separation of the high-energy and lower-divergence (collimated) protons with a quasi-monoenergetic spectrum produced by the shock. The resulting transverse spectral filtering of protons is demonstrated in Fig.~\ref{3D_defoc}(c), where we show the proton 3D trace space ($z-\theta_x-\theta_y$), and its 2D projection ($\theta_x-\theta_y$) after exiting the plasma. For this figure, we have selected only particles with energies $\epsilon_\text{prot}\geq 80$ MeV, and have divided them into the collimated $p_{x,y}/p_z<100$~mrad (orange) and divergent $p_{x,y}/p_z>100$~mrad (green) parts. The divergent protons form a ring in the angular space, while collimated particles remain on-axis. The spectra of these two groups are shown by the corresponding colors in Fig.~\ref{3D_defoc}(d), and we see that the collimated protons form a quasi-monoenergetic beam peaked at 240 MeV with the maximum energy close to 330 MeV, and a narrow spread (FWHM) of $20\%$ (orange in Fig.~\ref{3D_defoc}(d)). This beam has a small divergence ($\simeq 1.5^\circ$ half angle), and for the case of a 800 nm laser, the total particle number of this beam estimates as $6\times10^9$.

\begin{figure}[!ht]
  \centering
  \includegraphics[width=\linewidth]{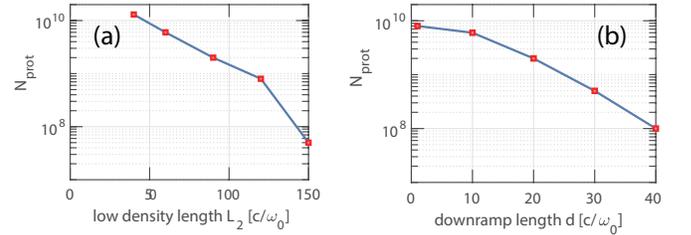}\\
  \caption{\label{parameter_scan} The relations between numbers of collimated quasimonoenergetic protons $N_{prot}$ ($\theta_{x,y}<100$ mrad) and the lengths of low density region $L_2$ (a) and density decreasing region $d$ (b) by fixing all other parameters.}
\end{figure}

The acceleration and defocusing processes are determined by the shock dynamics. It was shown previously, that the plasma Mach number for the collisionless electrostatic shock is $M_\text{sh}\simeq1-2$, and shock formation time is roughly $t_\text{sh} \simeq 4\pi/\omega_\text{pi}$,  where $\omega_\text{pi}$ is the proton plasma frequency of high density region \cite{forslund1970csa,fiuza2013pop}. While it is being formed, the shock propagates a distance $L_\text{sh} = v_\text{sh} t_\text{sh}$, and therefore, the plasma length of the second stage needs to be larger than this distance, $d+L_2\geq L_\text{sh} \simeq 60\ c/\omega_0$. To understand better the role of plasma length, we have performed the parametric studies, by taking the total number of the collimated protons ($\theta_{x,y}<100$ mrad) as a figure of merit. In Fig.~\ref{parameter_scan}(a), we show the scan of $L_2$, which demonstrates that for the longer $L_2$, the total charge falls as the on-axis protons become more divergent. On the other hand, for the shorter $L_2$ the total charge increases, but so does the proton energy spread, and for $L_2=20\ c/\omega_0$, their spectrum becomes continuous (not shown in the figure).

We find that the efficiency of the shock reflection can be controlled by varying the length of the downramp plasma section. In Fig.~\ref{parameter_scan}(b), we confirm this by changing the downramp length for a fixed total length $d+L_2=L_\text{sh}$. One can see that total quasi-mononenergetic charge increases for shorter ramps, and this growth saturates for $d\lesssim 10\,c/\omega_0$. From the performed analysis, we obtain a roughly estimate of the optimal scaling [$d\leq L_\text{sh}/2$, $L_{sh}/2\leq L_2\leq 2L_{sh}$], which assures production of $\sim 10^9$ quasi-monoenergetic protons.

Let us now derive a simple estimate of the maximal proton energy, which has a contribution from both the wakefield and shock stages. The maximal velocity achieved can be expressed as ${v_\text{max} = (v_\text{csa}+v_\text{wf}) / (1+v_\text{csa}v_\text{wf}/c^2)}$, where ${v_\text{csa}=(2v_\text{sh})/(1+v_\text{sh}^2/c^2)}$ is the velocity gained from the shock, and $v_{wf}$ is obtained from the NCD wakefield. Assuming $v_\text{wf}\ll c$,
the wakefield contribution can be estimated as $v_\text{wf} \approx e\psi_\text{max}/m_p$, where $\psi_\text{max}$ is the maximum potential of the longitudinal electric field. For shock formation with a sharp density ramp we consider no magnetic vortex pressure \cite{bulanov2005,nakamura2010}, and we also assume uniform shock propagation, $v_{sh}=M_{sh}c_s$, where $c_s=\sqrt{k_BT_h/m_p}$ is the proton acoustic velocity, $T_h$ is the temperature of hot plasma electrons, and $k_B$ is the Boltzmann constant. Expanding $v_\text{max}$ as a series of $v_\text{wf}$, we can estimate the maximum proton energy as:
\begin{eqnarray}\label{1D_ion_energy}
\epsilon_\text{max}\simeq 2M_\text{sh}^2 k_BT_h+2M_\text{sh}e\psi_\text{max}\sqrt{\frac{k_BT_h}{m_p}}
+\frac{e^2\psi_\text{max}^2}{2m_p}\,.
\end{eqnarray}

In case of a laser pulse matched to the plasma density, electrons are completely blown out from the wake, and $\psi_\text{max} \propto a_0$ \cite{lu2006, pukhov2002}. In order to estimate $T_h$, we may consider balance between the plasma electron energy area density and the absorbed laser energy area density \cite{wan2018tri}. Assuming a uniform temperature in NCD plasma region balance equation reads ${3k_B T_h\;L_1 n_e=\eta I\tau_\text{laser}}$, where $I$ is laser intensity, $\tau_\text{laser}$ is the laser pulse duration, and $\eta$ is the laser to electron conversion efficiency. Therefore, knowing the laser absorption efficiency $\eta$, and coefficient between $\psi_\text{max}$ and $a_0$, one can readily get the maximum proton energy.

\begin{figure}[!ht]
  \centering
  \includegraphics[width=\linewidth]{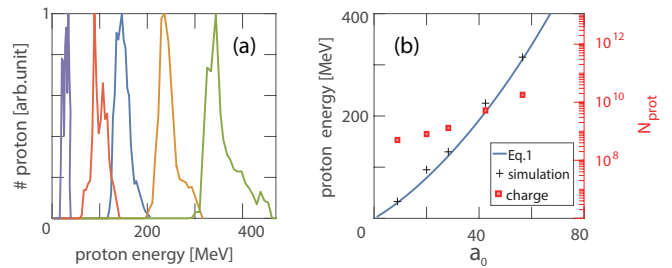}\\
  \caption{\label{3D_ion_scaling}(a) Spectrum of proton beams for different laser plasma parameters corresponding to $a_0$ = 9 (purple), $20$ (red), $28$ (blue), $42$ (orange) and $55$ (green). (b) The relation between proton peak energy $\epsilon_\text{max}$ and particle numbers $N_\text{prot}$ with laser $a_0$}.
\end{figure}

In order to verify this scaling, and to find the proper coefficients, we have performed another parametric study. In this study we have scaled up the laser intensity and plasma densities, but fixed the laser and plasma spatial scales (laser spot size/duration, plasma $L$ and $d$ lengths) to be the same as the simulation shown in Fig.~\ref{3Dexample_stages} and Fig.~\ref{3D_defoc}. We have scanned $a_0$ from $9$ to $55$, at the same time increasing $n_1$ proportionally from $0.6\, n_c$ to $4\, n_c$, to maintain the matched spot size. The ratio $n_1/n_2$ was chosen as $10$, for effective shock formation \cite{fiuza2013pop}. From these simulations we could estimate the coupling efficiency $\eta=30\%$ ($k_BT_h/m_ec^2\sim a_0$), and $e\psi_\text{max}/m_ec^2\sim 4 a_0$. In Fig.~\ref{3D_ion_scaling}(a) we demonstrate the proton energy spectra for the different $a_0$, which shows that the quasi-monoenergetic feature is preserved reasonably along the scaling. The scaling for the proton peak energy is plotted in Fig.~\ref{3D_ion_scaling}(b), and it agrees well with the theory Eq.~\ref{1D_ion_energy}, which indicates that at low laser intensities (e.g. $a_0=9$) acceleration is mainly dominated by shock acceleration (1st term of Eq.~\ref{1D_ion_energy}) and scales as $\epsilon_\text{max}\propto a_0$, while for higher intensities, the wakefield contribution becomes significant (2nd and 3rd terms of Eq.~\ref{1D_ion_energy}), which improves the scaling to $\propto a_0^{3/2}$ and further to $\propto a_0^2$. The particle numbers in all cases are around $10^{9}-10^{10}$, and their divergence angles decrease with the increased laser power from 3$^\circ$ to 1.2$^\circ$.

The proposed scheme can be readily tested with ultraintense ultrashort lasers, and our model predicts that high quality proton beams with energies from 100 to 300 MeV can be obtained using laser powers from 200 TW to 1 PW ($a_0=20-42$). One option can be a commercially available Ti:Sapphire laser system with $\lambda_0=800$~nm wavelength, which requires a few tens of microns long NCD plasma ($\sim 10^{20-21}$ cm$^{-3}$) with a downramp of a few micrometers. Such a target can be fabricated using a high pressure gas jet \cite{sylla2012}, with a density ramp formed by laser machining of the plasma. For this machining a low-power laser is used to pre-heat plasma locally, thus creating small-scale features in its density profile \cite{Pai2005}. Another option can be to use a CO$_2$ laser system with $\lambda_0=10\mu$m, for which NCD can be achieved with gas densities of $\sim 10^{19}$ cm$^{-3}$ \cite{haberberger2012}. The next generations of CO$_2$ lasers promise to deliver short 100 TW pulses \cite{pogorelsky2014} that are capable of high quality beams of 100 MeV-level protons.

In summary, we have presented a new concept for efficiently producing quasi-monoenergetic proton beams from NCD plasma with compact lasers. In the proposed scheme, a sharply tailored density plasma is used to generate a wake and a small size dense plasma shock which accelerates protons and triggers a defocusing process to truncate the particles phase separating a high energy collimated beam.

This work was supported by NSFC Grant No. 11425521, No. 11535006, No. 11475101 and No. 11775125, the Thousand Young Talents Program. Simulations were performed on Sunway TaihuLight cluster at National Supercomputing Center and Edison cluster at NERSC.

%\bibliography{tailored_plasma_refs}

\begin{thebibliography}{42}%
\makeatletter
\providecommand \@ifxundefined [1]{%
 \@ifx{#1\undefined}
}%
\providecommand \@ifnum [1]{%
 \ifnum #1\expandafter \@firstoftwo
 \else \expandafter \@secondoftwo
 \fi
}%
\providecommand \@ifx [1]{%
 \ifx #1\expandafter \@firstoftwo
 \else \expandafter \@secondoftwo
 \fi
}%
\providecommand \natexlab [1]{#1}%
\providecommand \enquote  [1]{``#1''}%
\providecommand \bibnamefont  [1]{#1}%
\providecommand \bibfnamefont [1]{#1}%
\providecommand \citenamefont [1]{#1}%
\providecommand \href@noop [0]{\@secondoftwo}%
\providecommand \href [0]{\begingroup \@sanitize@url \@href}%
\providecommand \@href[1]{\@@startlink{#1}\@@href}%
\providecommand \@@href[1]{\endgroup#1\@@endlink}%
\providecommand \@sanitize@url [0]{\catcode `\\12\catcode `\$12\catcode
  `\&12\catcode `\#12\catcode `\^12\catcode `\_12\catcode `\%12\relax}%
\providecommand \@@startlink[1]{}%
\providecommand \@@endlink[0]{}%
\providecommand \url  [0]{\begingroup\@sanitize@url \@url }%
\providecommand \@url [1]{\endgroup\@href {#1}{\urlprefix }}%
\providecommand \urlprefix  [0]{URL }%
\providecommand \Eprint [0]{\href }%
\providecommand \doibase [0]{http://dx.doi.org/}%
\providecommand \selectlanguage [0]{\@gobble}%
\providecommand \bibinfo  [0]{\@secondoftwo}%
\providecommand \bibfield  [0]{\@secondoftwo}%
\providecommand \translation [1]{[#1]}%
\providecommand \BibitemOpen [0]{}%
\providecommand \bibitemStop [0]{}%
\providecommand \bibitemNoStop [0]{.\EOS\space}%
\providecommand \EOS [0]{\spacefactor3000\relax}%
\providecommand \BibitemShut  [1]{\csname bibitem#1\endcsname}%
\let\auto@bib@innerbib\@empty
%</preamble>
\bibitem [{\citenamefont {Mourou}\ \emph {et~al.}(2006)\citenamefont {Mourou},
  \citenamefont {Tajima},\ and\ \citenamefont {Bulanov}}]{mourou2006optics}%
  \BibitemOpen
  \bibfield  {author} {\bibinfo {author} {\bibfnamefont {G.~A.}\ \bibnamefont
  {Mourou}}, \bibinfo {author} {\bibfnamefont {T.}~\bibnamefont {Tajima}}, \
  and\ \bibinfo {author} {\bibfnamefont {S.~V.}\ \bibnamefont {Bulanov}},\
  }\href@noop {} {\bibfield  {journal} {\bibinfo  {journal} {Reviews of modern
  physics}\ }\textbf {\bibinfo {volume} {78}},\ \bibinfo {pages} {309}
  (\bibinfo {year} {2006})}\BibitemShut {NoStop}%
\bibitem [{\citenamefont {Daido}\ \emph {et~al.}(2012)\citenamefont {Daido},
  \citenamefont {Nishiuchi},\ and\ \citenamefont {Pirozhkov}}]{daido2012}%
  \BibitemOpen
  \bibfield  {author} {\bibinfo {author} {\bibfnamefont {H.}~\bibnamefont
  {Daido}}, \bibinfo {author} {\bibfnamefont {M.}~\bibnamefont {Nishiuchi}}, \
  and\ \bibinfo {author} {\bibfnamefont {A.~S.}\ \bibnamefont {Pirozhkov}},\
  }\href@noop {} {\bibfield  {journal} {\bibinfo  {journal} {Reports on
  Progress in Physics}\ }\textbf {\bibinfo {volume} {75}},\ \bibinfo {pages}
  {056401} (\bibinfo {year} {2012})}\BibitemShut {NoStop}%
\bibitem [{\citenamefont {Macchi}\ \emph {et~al.}(2013)\citenamefont {Macchi},
  \citenamefont {Borghesi},\ and\ \citenamefont {Passoni}}]{macchi2013}%
  \BibitemOpen
  \bibfield  {author} {\bibinfo {author} {\bibfnamefont {A.}~\bibnamefont
  {Macchi}}, \bibinfo {author} {\bibfnamefont {M.}~\bibnamefont {Borghesi}}, \
  and\ \bibinfo {author} {\bibfnamefont {M.}~\bibnamefont {Passoni}},\
  }\href@noop {} {\bibfield  {journal} {\bibinfo  {journal} {Reviews of Modern
  Physics}\ }\textbf {\bibinfo {volume} {85}},\ \bibinfo {pages} {751}
  (\bibinfo {year} {2013})}\BibitemShut {NoStop}%
\bibitem [{\citenamefont {Borghesi}\ \emph {et~al.}(2002)\citenamefont
  {Borghesi}, \citenamefont {Bulanov}, \citenamefont {Campbell}, \citenamefont
  {Clarke}, \citenamefont {Esirkepov}, \citenamefont {Galimberti},
  \citenamefont {Gizzi}, \citenamefont {MacKinnon}, \citenamefont {Naumova},
  \citenamefont {Pegoraro}, \citenamefont {Ruhl}, \citenamefont {Schiavi},\
  and\ \citenamefont {Willi}}]{borghesi2002PIprl}%
  \BibitemOpen
  \bibfield  {author} {\bibinfo {author} {\bibfnamefont {M.}~\bibnamefont
  {Borghesi}}, \bibinfo {author} {\bibfnamefont {S.}~\bibnamefont {Bulanov}},
  \bibinfo {author} {\bibfnamefont {D.~H.}\ \bibnamefont {Campbell}}, \bibinfo
  {author} {\bibfnamefont {R.~J.}\ \bibnamefont {Clarke}}, \bibinfo {author}
  {\bibfnamefont {T.~Z.}\ \bibnamefont {Esirkepov}}, \bibinfo {author}
  {\bibfnamefont {M.}~\bibnamefont {Galimberti}}, \bibinfo {author}
  {\bibfnamefont {L.~A.}\ \bibnamefont {Gizzi}}, \bibinfo {author}
  {\bibfnamefont {A.~J.}\ \bibnamefont {MacKinnon}}, \bibinfo {author}
  {\bibfnamefont {N.~M.}\ \bibnamefont {Naumova}}, \bibinfo {author}
  {\bibfnamefont {F.}~\bibnamefont {Pegoraro}}, \bibinfo {author}
  {\bibfnamefont {H.}~\bibnamefont {Ruhl}}, \bibinfo {author} {\bibfnamefont
  {A.}~\bibnamefont {Schiavi}}, \ and\ \bibinfo {author} {\bibfnamefont
  {O.}~\bibnamefont {Willi}},\ }\href@noop {} {\bibfield  {journal} {\bibinfo
  {journal} {Physical Review Letters}\ }\textbf {\bibinfo {volume} {88}},\
  \bibinfo {pages} {135002} (\bibinfo {year} {2002})}\BibitemShut {NoStop}%
\bibitem [{\citenamefont {Li}\ \emph {et~al.}(2006)\citenamefont {Li},
  \citenamefont {Seguin}, \citenamefont {Frenje}, \citenamefont {Rygg},
  \citenamefont {Petrasso}, \citenamefont {Town}, \citenamefont {Amendt},
  \citenamefont {Hatchett}, \citenamefont {Landen}, \citenamefont {Mackinnon},
  \citenamefont {Patel}, \citenamefont {Smalyuk}, \citenamefont {Sangster},\
  and\ \citenamefont {Knauer}}]{li2006pi}%
  \BibitemOpen
  \bibfield  {author} {\bibinfo {author} {\bibfnamefont {C.~K.}\ \bibnamefont
  {Li}}, \bibinfo {author} {\bibfnamefont {F.~H.}\ \bibnamefont {Seguin}},
  \bibinfo {author} {\bibfnamefont {J.~A.}\ \bibnamefont {Frenje}}, \bibinfo
  {author} {\bibfnamefont {J.~R.}\ \bibnamefont {Rygg}}, \bibinfo {author}
  {\bibfnamefont {R.~D.}\ \bibnamefont {Petrasso}}, \bibinfo {author}
  {\bibfnamefont {R.~P.~J.}\ \bibnamefont {Town}}, \bibinfo {author}
  {\bibfnamefont {P.~A.}\ \bibnamefont {Amendt}}, \bibinfo {author}
  {\bibfnamefont {S.~P.}\ \bibnamefont {Hatchett}}, \bibinfo {author}
  {\bibfnamefont {O.~L.}\ \bibnamefont {Landen}}, \bibinfo {author}
  {\bibfnamefont {A.~J.}\ \bibnamefont {Mackinnon}}, \bibinfo {author}
  {\bibfnamefont {P.~K.}\ \bibnamefont {Patel}}, \bibinfo {author}
  {\bibfnamefont {V.~A.}\ \bibnamefont {Smalyuk}}, \bibinfo {author}
  {\bibfnamefont {T.~C.}\ \bibnamefont {Sangster}}, \ and\ \bibinfo {author}
  {\bibfnamefont {J.~P.}\ \bibnamefont {Knauer}},\ }\href@noop {} {\bibfield
  {journal} {\bibinfo  {journal} {Physical Review Letters}\ }\textbf {\bibinfo
  {volume} {97}},\ \bibinfo {pages} {135003} (\bibinfo {year}
  {2006})}\BibitemShut {NoStop}%
\bibitem [{\citenamefont {Bulanov}\ \emph {et~al.}(2002)\citenamefont
  {Bulanov}, \citenamefont {Esirkepov}, \citenamefont {Khoroshkov},
  \citenamefont {Kunetsov},\ and\ \citenamefont {Pegoraro}}]{bulanov2002CTpra}%
  \BibitemOpen
  \bibfield  {author} {\bibinfo {author} {\bibfnamefont {S.~V.}\ \bibnamefont
  {Bulanov}}, \bibinfo {author} {\bibfnamefont {T.~Z.}\ \bibnamefont
  {Esirkepov}}, \bibinfo {author} {\bibfnamefont {V.~S.}\ \bibnamefont
  {Khoroshkov}}, \bibinfo {author} {\bibfnamefont {A.~V.}\ \bibnamefont
  {Kunetsov}}, \ and\ \bibinfo {author} {\bibfnamefont {F.}~\bibnamefont
  {Pegoraro}},\ }\href@noop {} {\bibfield  {journal} {\bibinfo  {journal}
  {Physics Letters A}\ }\textbf {\bibinfo {volume} {299}},\ \bibinfo {pages}
  {240} (\bibinfo {year} {2002})}\BibitemShut {NoStop}%
\bibitem [{\citenamefont {Malka}\ \emph {et~al.}(2004)\citenamefont {Malka},
  \citenamefont {Fritzler}, \citenamefont {Lefebvre}, \citenamefont
  {d'Humieres}, \citenamefont {Ferrand}, \citenamefont {Grillon}, \citenamefont
  {Albaret}, \citenamefont {Meyroneinc}, \citenamefont {Chambaret},
  \citenamefont {Antonetti},\ and\ \citenamefont {Hulin}}]{malka2004ct}%
  \BibitemOpen
  \bibfield  {author} {\bibinfo {author} {\bibfnamefont {V.}~\bibnamefont
  {Malka}}, \bibinfo {author} {\bibfnamefont {S.}~\bibnamefont {Fritzler}},
  \bibinfo {author} {\bibfnamefont {E.}~\bibnamefont {Lefebvre}}, \bibinfo
  {author} {\bibfnamefont {E.}~\bibnamefont {d'Humieres}}, \bibinfo {author}
  {\bibfnamefont {R.}~\bibnamefont {Ferrand}}, \bibinfo {author} {\bibfnamefont
  {G.}~\bibnamefont {Grillon}}, \bibinfo {author} {\bibfnamefont
  {C.}~\bibnamefont {Albaret}}, \bibinfo {author} {\bibfnamefont
  {S.}~\bibnamefont {Meyroneinc}}, \bibinfo {author} {\bibfnamefont {J.~P.}\
  \bibnamefont {Chambaret}}, \bibinfo {author} {\bibfnamefont {A.}~\bibnamefont
  {Antonetti}}, \ and\ \bibinfo {author} {\bibfnamefont {D.}~\bibnamefont
  {Hulin}},\ }\href@noop {} {\bibfield  {journal} {\bibinfo  {journal} {Medical
  Physics}\ }\textbf {\bibinfo {volume} {31}},\ \bibinfo {pages} {1587}
  (\bibinfo {year} {2004})}\BibitemShut {NoStop}%
\bibitem [{\citenamefont {Snavely}\ \emph {et~al.}(2000)\citenamefont
  {Snavely}, \citenamefont {Key}, \citenamefont {Hatchett}, \citenamefont
  {Cowan}, \citenamefont {Roth}, \citenamefont {Phillips}, \citenamefont
  {Stoyer}, \citenamefont {Henry}, \citenamefont {Sangster}, \citenamefont
  {Singh}, \citenamefont {Wilks}, \citenamefont {MacKinnon}, \citenamefont
  {Offenberger}, \citenamefont {Pennington}, \citenamefont {Yasuike},
  \citenamefont {Langdon}, \citenamefont {Lasinski}, \citenamefont {Johnson},
  \citenamefont {Perry},\ and\ \citenamefont {Campbell}}]{snavely2000TNSA}%
  \BibitemOpen
  \bibfield  {author} {\bibinfo {author} {\bibfnamefont {R.~A.}\ \bibnamefont
  {Snavely}}, \bibinfo {author} {\bibfnamefont {M.~H.}\ \bibnamefont {Key}},
  \bibinfo {author} {\bibfnamefont {S.~P.}\ \bibnamefont {Hatchett}}, \bibinfo
  {author} {\bibfnamefont {T.~E.}\ \bibnamefont {Cowan}}, \bibinfo {author}
  {\bibfnamefont {M.}~\bibnamefont {Roth}}, \bibinfo {author} {\bibfnamefont
  {T.~W.}\ \bibnamefont {Phillips}}, \bibinfo {author} {\bibfnamefont {M.~A.}\
  \bibnamefont {Stoyer}}, \bibinfo {author} {\bibfnamefont {E.~A.}\
  \bibnamefont {Henry}}, \bibinfo {author} {\bibfnamefont {T.~C.}\ \bibnamefont
  {Sangster}}, \bibinfo {author} {\bibfnamefont {M.~S.}\ \bibnamefont {Singh}},
  \bibinfo {author} {\bibfnamefont {S.~C.}\ \bibnamefont {Wilks}}, \bibinfo
  {author} {\bibfnamefont {A.}~\bibnamefont {MacKinnon}}, \bibinfo {author}
  {\bibfnamefont {A.}~\bibnamefont {Offenberger}}, \bibinfo {author}
  {\bibfnamefont {D.~M.}\ \bibnamefont {Pennington}}, \bibinfo {author}
  {\bibfnamefont {K.}~\bibnamefont {Yasuike}}, \bibinfo {author} {\bibfnamefont
  {A.~B.}\ \bibnamefont {Langdon}}, \bibinfo {author} {\bibfnamefont {B.~F.}\
  \bibnamefont {Lasinski}}, \bibinfo {author} {\bibfnamefont {J.}~\bibnamefont
  {Johnson}}, \bibinfo {author} {\bibfnamefont {M.~D.}\ \bibnamefont {Perry}},
  \ and\ \bibinfo {author} {\bibfnamefont {E.~M.}\ \bibnamefont {Campbell}},\
  }\href@noop {} {\bibfield  {journal} {\bibinfo  {journal} {Physical Review
  Letters}\ }\textbf {\bibinfo {volume} {85}},\ \bibinfo {pages} {2945}
  (\bibinfo {year} {2000})}\BibitemShut {NoStop}%
\bibitem [{\citenamefont {Esirkepov}\ \emph {et~al.}(2004)\citenamefont
  {Esirkepov}, \citenamefont {Borghesi}, \citenamefont {Bulanov}, \citenamefont
  {Mourou},\ and\ \citenamefont {Tajima}}]{esirkepov2004}%
  \BibitemOpen
  \bibfield  {author} {\bibinfo {author} {\bibfnamefont {T.}~\bibnamefont
  {Esirkepov}}, \bibinfo {author} {\bibfnamefont {M.}~\bibnamefont {Borghesi}},
  \bibinfo {author} {\bibfnamefont {S.~V.}\ \bibnamefont {Bulanov}}, \bibinfo
  {author} {\bibfnamefont {G.}~\bibnamefont {Mourou}}, \ and\ \bibinfo {author}
  {\bibfnamefont {T.}~\bibnamefont {Tajima}},\ }\href@noop {} {\bibfield
  {journal} {\bibinfo  {journal} {Physical Review Letters}\ }\textbf {\bibinfo
  {volume} {92}},\ \bibinfo {pages} {175003} (\bibinfo {year}
  {2004})}\BibitemShut {NoStop}%
\bibitem [{\citenamefont {Macchi}\ \emph {et~al.}(2005)\citenamefont {Macchi},
  \citenamefont {Cattani}, \citenamefont {Liseykina},\ and\ \citenamefont
  {Cornolti}}]{macchi2005}%
  \BibitemOpen
  \bibfield  {author} {\bibinfo {author} {\bibfnamefont {A.}~\bibnamefont
  {Macchi}}, \bibinfo {author} {\bibfnamefont {F.}~\bibnamefont {Cattani}},
  \bibinfo {author} {\bibfnamefont {T.~V.}\ \bibnamefont {Liseykina}}, \ and\
  \bibinfo {author} {\bibfnamefont {F.}~\bibnamefont {Cornolti}},\ }\href@noop
  {} {\bibfield  {journal} {\bibinfo  {journal} {Physical review letters}\
  }\textbf {\bibinfo {volume} {94}},\ \bibinfo {pages} {165003} (\bibinfo
  {year} {2005})}\BibitemShut {NoStop}%
\bibitem [{\citenamefont {Zhang}\ \emph {et~al.}(2007)\citenamefont {Zhang},
  \citenamefont {Shen}, \citenamefont {Li}, \citenamefont {Jin},\ and\
  \citenamefont {Wang}}]{zhang2007a}%
  \BibitemOpen
  \bibfield  {author} {\bibinfo {author} {\bibfnamefont {X.~M.}\ \bibnamefont
  {Zhang}}, \bibinfo {author} {\bibfnamefont {B.~F.}\ \bibnamefont {Shen}},
  \bibinfo {author} {\bibfnamefont {X.~M.}\ \bibnamefont {Li}}, \bibinfo
  {author} {\bibfnamefont {Z.~Y.}\ \bibnamefont {Jin}}, \ and\ \bibinfo
  {author} {\bibfnamefont {F.~C.}\ \bibnamefont {Wang}},\ }\href@noop {}
  {\bibfield  {journal} {\bibinfo  {journal} {Physics of Plasmas}\ }\textbf
  {\bibinfo {volume} {14}},\ \bibinfo {pages} {073101} (\bibinfo {year}
  {2007})}\BibitemShut {NoStop}%
\bibitem [{\citenamefont {Robinson}\ \emph {et~al.}(2008)\citenamefont
  {Robinson}, \citenamefont {Zepf}, \citenamefont {Kar}, \citenamefont
  {Evans},\ and\ \citenamefont {Bellei}}]{robinson2008}%
  \BibitemOpen
  \bibfield  {author} {\bibinfo {author} {\bibfnamefont {A.~P.~L.}\
  \bibnamefont {Robinson}}, \bibinfo {author} {\bibfnamefont {M.}~\bibnamefont
  {Zepf}}, \bibinfo {author} {\bibfnamefont {S.}~\bibnamefont {Kar}}, \bibinfo
  {author} {\bibfnamefont {R.~G.}\ \bibnamefont {Evans}}, \ and\ \bibinfo
  {author} {\bibfnamefont {C.}~\bibnamefont {Bellei}},\ }\href@noop {}
  {\bibfield  {journal} {\bibinfo  {journal} {New Journal of Physics}\ }\textbf
  {\bibinfo {volume} {10}},\ \bibinfo {pages} {013021} (\bibinfo {year}
  {2008})}\BibitemShut {NoStop}%
\bibitem [{\citenamefont {Klimo}\ \emph {et~al.}(2008)\citenamefont {Klimo},
  \citenamefont {Psikal}, \citenamefont {Limpouch},\ and\ \citenamefont
  {Tikhonchuk}}]{klimo2008}%
  \BibitemOpen
  \bibfield  {author} {\bibinfo {author} {\bibfnamefont {O.}~\bibnamefont
  {Klimo}}, \bibinfo {author} {\bibfnamefont {J.}~\bibnamefont {Psikal}},
  \bibinfo {author} {\bibfnamefont {J.}~\bibnamefont {Limpouch}}, \ and\
  \bibinfo {author} {\bibfnamefont {V.~T.}\ \bibnamefont {Tikhonchuk}},\
  }\href@noop {} {\bibfield  {journal} {\bibinfo  {journal} {Physical Review
  Special Topics-Accelerators and Beams}\ }\textbf {\bibinfo {volume} {11}},\
  \bibinfo {pages} {031301} (\bibinfo {year} {2008})}\BibitemShut {NoStop}%
\bibitem [{\citenamefont {Yan}\ \emph {et~al.}(2008)\citenamefont {Yan},
  \citenamefont {Lin}, \citenamefont {Sheng}, \citenamefont {Guo},
  \citenamefont {Liu}, \citenamefont {Lu}, \citenamefont {Fang},\ and\
  \citenamefont {Chen}}]{yan2008}%
  \BibitemOpen
  \bibfield  {author} {\bibinfo {author} {\bibfnamefont {X.~Q.}\ \bibnamefont
  {Yan}}, \bibinfo {author} {\bibfnamefont {C.}~\bibnamefont {Lin}}, \bibinfo
  {author} {\bibfnamefont {Z.~M.}\ \bibnamefont {Sheng}}, \bibinfo {author}
  {\bibfnamefont {Z.~Y.}\ \bibnamefont {Guo}}, \bibinfo {author} {\bibfnamefont
  {B.~C.}\ \bibnamefont {Liu}}, \bibinfo {author} {\bibfnamefont {Y.~R.}\
  \bibnamefont {Lu}}, \bibinfo {author} {\bibfnamefont {J.~X.}\ \bibnamefont
  {Fang}}, \ and\ \bibinfo {author} {\bibfnamefont {J.~E.}\ \bibnamefont
  {Chen}},\ }\href@noop {} {\bibfield  {journal} {\bibinfo  {journal} {Physical
  Review Letters}\ }\textbf {\bibinfo {volume} {100}},\ \bibinfo {pages}
  {135003} (\bibinfo {year} {2008})}\BibitemShut {NoStop}%
\bibitem [{\citenamefont {Wan}\ \emph {et~al.}(2016)\citenamefont {Wan},
  \citenamefont {Pai}, \citenamefont {Zhang}, \citenamefont {Li}, \citenamefont
  {Wu}, \citenamefont {Hua}, \citenamefont {Lu}, \citenamefont {Gu},
  \citenamefont {Silva}, \citenamefont {Joshi},\ and\ \citenamefont
  {Mori}}]{wan2016}%
  \BibitemOpen
  \bibfield  {author} {\bibinfo {author} {\bibfnamefont {Y.}~\bibnamefont
  {Wan}}, \bibinfo {author} {\bibfnamefont {C.~H.}\ \bibnamefont {Pai}},
  \bibinfo {author} {\bibfnamefont {C.~J.}\ \bibnamefont {Zhang}}, \bibinfo
  {author} {\bibfnamefont {F.}~\bibnamefont {Li}}, \bibinfo {author}
  {\bibfnamefont {Y.~P.}\ \bibnamefont {Wu}}, \bibinfo {author} {\bibfnamefont
  {J.~F.}\ \bibnamefont {Hua}}, \bibinfo {author} {\bibfnamefont
  {W.}~\bibnamefont {Lu}}, \bibinfo {author} {\bibfnamefont {Y.~Q.}\
  \bibnamefont {Gu}}, \bibinfo {author} {\bibfnamefont {L.~O.}\ \bibnamefont
  {Silva}}, \bibinfo {author} {\bibfnamefont {C.}~\bibnamefont {Joshi}}, \ and\
  \bibinfo {author} {\bibfnamefont {W.~B.}\ \bibnamefont {Mori}},\ }\href@noop
  {} {\bibfield  {journal} {\bibinfo  {journal} {Physical Review Letters}\
  }\textbf {\bibinfo {volume} {117}},\ \bibinfo {pages} {234801} (\bibinfo
  {year} {2016})}\BibitemShut {NoStop}%
\bibitem [{\citenamefont {Albright}\ \emph {et~al.}(2007)\citenamefont
  {Albright}, \citenamefont {Yin}, \citenamefont {Bowers}, \citenamefont
  {Hegelich}, \citenamefont {Flippo}, \citenamefont {Kwan},\ and\ \citenamefont
  {Fernandez}}]{albright2007boa}%
  \BibitemOpen
  \bibfield  {author} {\bibinfo {author} {\bibfnamefont {B.~J.}\ \bibnamefont
  {Albright}}, \bibinfo {author} {\bibfnamefont {L.}~\bibnamefont {Yin}},
  \bibinfo {author} {\bibfnamefont {K.~J.}\ \bibnamefont {Bowers}}, \bibinfo
  {author} {\bibfnamefont {B.~M.}\ \bibnamefont {Hegelich}}, \bibinfo {author}
  {\bibfnamefont {K.~A.}\ \bibnamefont {Flippo}}, \bibinfo {author}
  {\bibfnamefont {T.~J.~T.}\ \bibnamefont {Kwan}}, \ and\ \bibinfo {author}
  {\bibfnamefont {J.~C.}\ \bibnamefont {Fernandez}},\ }\href@noop {} {\bibfield
   {journal} {\bibinfo  {journal} {Physics of Plasmas}\ }\textbf {\bibinfo
  {volume} {14}},\ \bibinfo {pages} {094502} (\bibinfo {year}
  {2007})}\BibitemShut {NoStop}%
\bibitem [{\citenamefont {Yin}\ \emph {et~al.}(2007)\citenamefont {Yin},
  \citenamefont {Albright}, \citenamefont {Hegelich}, \citenamefont {Bowers},
  \citenamefont {Flippo}, \citenamefont {Kwan},\ and\ \citenamefont
  {Fernandez}}]{yin2007}%
  \BibitemOpen
  \bibfield  {author} {\bibinfo {author} {\bibfnamefont {L.}~\bibnamefont
  {Yin}}, \bibinfo {author} {\bibfnamefont {B.~J.}\ \bibnamefont {Albright}},
  \bibinfo {author} {\bibfnamefont {B.~M.}\ \bibnamefont {Hegelich}}, \bibinfo
  {author} {\bibfnamefont {K.~J.}\ \bibnamefont {Bowers}}, \bibinfo {author}
  {\bibfnamefont {K.~A.}\ \bibnamefont {Flippo}}, \bibinfo {author}
  {\bibfnamefont {T.~J.~T.}\ \bibnamefont {Kwan}}, \ and\ \bibinfo {author}
  {\bibfnamefont {J.~C.}\ \bibnamefont {Fernandez}},\ }\href@noop {} {\bibfield
   {journal} {\bibinfo  {journal} {Physics of Plasmas}\ }\textbf {\bibinfo
  {volume} {14}},\ \bibinfo {pages} {056706} (\bibinfo {year}
  {2007})}\BibitemShut {NoStop}%
\bibitem [{\citenamefont {Silva}\ \emph {et~al.}(2004)\citenamefont {Silva},
  \citenamefont {Marti}, \citenamefont {Davies}, \citenamefont {Fonseca},
  \citenamefont {Ren}, \citenamefont {Tsung},\ and\ \citenamefont
  {Mori}}]{silva2004}%
  \BibitemOpen
  \bibfield  {author} {\bibinfo {author} {\bibfnamefont {L.~O.}\ \bibnamefont
  {Silva}}, \bibinfo {author} {\bibfnamefont {M.}~\bibnamefont {Marti}},
  \bibinfo {author} {\bibfnamefont {J.~R.}\ \bibnamefont {Davies}}, \bibinfo
  {author} {\bibfnamefont {R.~A.}\ \bibnamefont {Fonseca}}, \bibinfo {author}
  {\bibfnamefont {C.}~\bibnamefont {Ren}}, \bibinfo {author} {\bibfnamefont
  {F.~S.}\ \bibnamefont {Tsung}}, \ and\ \bibinfo {author} {\bibfnamefont
  {W.~B.}\ \bibnamefont {Mori}},\ }\href@noop {} {\bibfield  {journal}
  {\bibinfo  {journal} {Physical Review Letters}\ }\textbf {\bibinfo {volume}
  {92}},\ \bibinfo {pages} {015002} (\bibinfo {year} {2004})}\BibitemShut
  {NoStop}%
\bibitem [{\citenamefont {Haberberger}\ \emph {et~al.}(2012)\citenamefont
  {Haberberger}, \citenamefont {Tochitsky}, \citenamefont {Fiuza},
  \citenamefont {Gong}, \citenamefont {Fonseca}, \citenamefont {Silva},
  \citenamefont {Mori},\ and\ \citenamefont {Joshi}}]{haberberger2012}%
  \BibitemOpen
  \bibfield  {author} {\bibinfo {author} {\bibfnamefont {D.}~\bibnamefont
  {Haberberger}}, \bibinfo {author} {\bibfnamefont {S.}~\bibnamefont
  {Tochitsky}}, \bibinfo {author} {\bibfnamefont {F.}~\bibnamefont {Fiuza}},
  \bibinfo {author} {\bibfnamefont {C.}~\bibnamefont {Gong}}, \bibinfo {author}
  {\bibfnamefont {R.~A.}\ \bibnamefont {Fonseca}}, \bibinfo {author}
  {\bibfnamefont {L.~O.}\ \bibnamefont {Silva}}, \bibinfo {author}
  {\bibfnamefont {W.~B.}\ \bibnamefont {Mori}}, \ and\ \bibinfo {author}
  {\bibfnamefont {C.}~\bibnamefont {Joshi}},\ }\href@noop {} {\bibfield
  {journal} {\bibinfo  {journal} {Nature Physics}\ }\textbf {\bibinfo {volume}
  {8}},\ \bibinfo {pages} {95} (\bibinfo {year} {2012})}\BibitemShut {NoStop}%
\bibitem [{\citenamefont {Fiuza}\ \emph {et~al.}(2012)\citenamefont {Fiuza},
  \citenamefont {Stockem}, \citenamefont {Boella}, \citenamefont {Fonseca},
  \citenamefont {Silva}, \citenamefont {Haberberger}, \citenamefont
  {Tochitsky}, \citenamefont {Gong}, \citenamefont {Mori},\ and\ \citenamefont
  {Joshi}}]{fiuza2012prl}%
  \BibitemOpen
  \bibfield  {author} {\bibinfo {author} {\bibfnamefont {F.}~\bibnamefont
  {Fiuza}}, \bibinfo {author} {\bibfnamefont {A.}~\bibnamefont {Stockem}},
  \bibinfo {author} {\bibfnamefont {E.}~\bibnamefont {Boella}}, \bibinfo
  {author} {\bibfnamefont {R.~A.}\ \bibnamefont {Fonseca}}, \bibinfo {author}
  {\bibfnamefont {L.~O.}\ \bibnamefont {Silva}}, \bibinfo {author}
  {\bibfnamefont {D.}~\bibnamefont {Haberberger}}, \bibinfo {author}
  {\bibfnamefont {S.}~\bibnamefont {Tochitsky}}, \bibinfo {author}
  {\bibfnamefont {C.}~\bibnamefont {Gong}}, \bibinfo {author} {\bibfnamefont
  {W.~B.}\ \bibnamefont {Mori}}, \ and\ \bibinfo {author} {\bibfnamefont
  {C.}~\bibnamefont {Joshi}},\ }\href@noop {} {\bibfield  {journal} {\bibinfo
  {journal} {Physical Review Letters}\ }\textbf {\bibinfo {volume} {109}},\
  \bibinfo {pages} {215001} (\bibinfo {year} {2012})}\BibitemShut {NoStop}%
\bibitem [{\citenamefont {Nakamura}\ \emph {et~al.}(2010)\citenamefont
  {Nakamura}, \citenamefont {Bulanov}, \citenamefont {Esirkepov},\ and\
  \citenamefont {Kando}}]{nakamura2010}%
  \BibitemOpen
  \bibfield  {author} {\bibinfo {author} {\bibfnamefont {T.}~\bibnamefont
  {Nakamura}}, \bibinfo {author} {\bibfnamefont {S.~V.}\ \bibnamefont
  {Bulanov}}, \bibinfo {author} {\bibfnamefont {T.~Z.}\ \bibnamefont
  {Esirkepov}}, \ and\ \bibinfo {author} {\bibfnamefont {M.}~\bibnamefont
  {Kando}},\ }\href@noop {} {\bibfield  {journal} {\bibinfo  {journal}
  {Physical Review Letters}\ }\textbf {\bibinfo {volume} {105}},\ \bibinfo
  {pages} {135002} (\bibinfo {year} {2010})}\BibitemShut {NoStop}%
\bibitem [{\citenamefont {Helle}\ \emph {et~al.}(2016)\citenamefont {Helle},
  \citenamefont {Gordon}, \citenamefont {Kaganovich}, \citenamefont {Chen},
  \citenamefont {Palastro},\ and\ \citenamefont {Ting}}]{helle2016}%
  \BibitemOpen
  \bibfield  {author} {\bibinfo {author} {\bibfnamefont {M.~H.}\ \bibnamefont
  {Helle}}, \bibinfo {author} {\bibfnamefont {D.~F.}\ \bibnamefont {Gordon}},
  \bibinfo {author} {\bibfnamefont {D.}~\bibnamefont {Kaganovich}}, \bibinfo
  {author} {\bibfnamefont {Y.}~\bibnamefont {Chen}}, \bibinfo {author}
  {\bibfnamefont {J.~P.}\ \bibnamefont {Palastro}}, \ and\ \bibinfo {author}
  {\bibfnamefont {A.}~\bibnamefont {Ting}},\ }\href@noop {} {\bibfield
  {journal} {\bibinfo  {journal} {Physical Review Letters}\ }\textbf {\bibinfo
  {volume} {117}},\ \bibinfo {pages} {165001} (\bibinfo {year}
  {2016})}\BibitemShut {NoStop}%
\bibitem [{\citenamefont {Kuznetsov}\ \emph {et~al.}(2001)\citenamefont
  {Kuznetsov}, \citenamefont {Esirkepov}, \citenamefont {Kamenets},\ and\
  \citenamefont {Bulanov}}]{kuznetsov2001}%
  \BibitemOpen
  \bibfield  {author} {\bibinfo {author} {\bibfnamefont {A.~V.}\ \bibnamefont
  {Kuznetsov}}, \bibinfo {author} {\bibfnamefont {T.~Z.}\ \bibnamefont
  {Esirkepov}}, \bibinfo {author} {\bibfnamefont {F.~F.}\ \bibnamefont
  {Kamenets}}, \ and\ \bibinfo {author} {\bibfnamefont {S.~V.}\ \bibnamefont
  {Bulanov}},\ }\href@noop {} {\bibfield  {journal} {\bibinfo  {journal}
  {Plasma Physics Reports}\ }\textbf {\bibinfo {volume} {27}},\ \bibinfo
  {pages} {211} (\bibinfo {year} {2001})}\BibitemShut {NoStop}%
\bibitem [{\citenamefont {Bulanov}\ \emph {et~al.}(2005)\citenamefont
  {Bulanov}, \citenamefont {Dylov}, \citenamefont {Esirkepov}, \citenamefont
  {Kamenets},\ and\ \citenamefont {Sokolov}}]{bulanov2005}%
  \BibitemOpen
  \bibfield  {author} {\bibinfo {author} {\bibfnamefont {S.~V.}\ \bibnamefont
  {Bulanov}}, \bibinfo {author} {\bibfnamefont {D.~V.}\ \bibnamefont {Dylov}},
  \bibinfo {author} {\bibfnamefont {T.~Z.}\ \bibnamefont {Esirkepov}}, \bibinfo
  {author} {\bibfnamefont {F.~F.}\ \bibnamefont {Kamenets}}, \ and\ \bibinfo
  {author} {\bibfnamefont {D.~V.}\ \bibnamefont {Sokolov}},\ }\href@noop {}
  {\bibfield  {journal} {\bibinfo  {journal} {Plasma Physics Reports}\ }\textbf
  {\bibinfo {volume} {31}},\ \bibinfo {pages} {369} (\bibinfo {year}
  {2005})}\BibitemShut {NoStop}%
\bibitem [{\citenamefont {Willingale}\ \emph {et~al.}(2006)\citenamefont
  {Willingale}, \citenamefont {Mangles}, \citenamefont {Nilson}, \citenamefont
  {Clarke}, \citenamefont {Dangor}, \citenamefont {Kaluza}, \citenamefont
  {Karsch}, \citenamefont {Lancaster}, \citenamefont {Mori}, \citenamefont
  {Najmudin}, \citenamefont {Schreiber}, \citenamefont {Thomas}, \citenamefont
  {Wei},\ and\ \citenamefont {Krushelnick}}]{willingale2006}%
  \BibitemOpen
  \bibfield  {author} {\bibinfo {author} {\bibfnamefont {L.}~\bibnamefont
  {Willingale}}, \bibinfo {author} {\bibfnamefont {S.~P.~D.}\ \bibnamefont
  {Mangles}}, \bibinfo {author} {\bibfnamefont {P.~M.}\ \bibnamefont {Nilson}},
  \bibinfo {author} {\bibfnamefont {R.~J.}\ \bibnamefont {Clarke}}, \bibinfo
  {author} {\bibfnamefont {A.~E.}\ \bibnamefont {Dangor}}, \bibinfo {author}
  {\bibfnamefont {M.~C.}\ \bibnamefont {Kaluza}}, \bibinfo {author}
  {\bibfnamefont {S.}~\bibnamefont {Karsch}}, \bibinfo {author} {\bibfnamefont
  {K.~L.}\ \bibnamefont {Lancaster}}, \bibinfo {author} {\bibfnamefont {W.~B.}\
  \bibnamefont {Mori}}, \bibinfo {author} {\bibfnamefont {Z.}~\bibnamefont
  {Najmudin}}, \bibinfo {author} {\bibfnamefont {J.}~\bibnamefont {Schreiber}},
  \bibinfo {author} {\bibfnamefont {A.~G.~R.}\ \bibnamefont {Thomas}}, \bibinfo
  {author} {\bibfnamefont {M.~S.}\ \bibnamefont {Wei}}, \ and\ \bibinfo
  {author} {\bibfnamefont {K.}~\bibnamefont {Krushelnick}},\ }\href@noop {}
  {\bibfield  {journal} {\bibinfo  {journal} {Physical Review Letters}\
  }\textbf {\bibinfo {volume} {96}},\ \bibinfo {pages} {245002} (\bibinfo
  {year} {2006})}\BibitemShut {NoStop}%
\bibitem [{\citenamefont {Yogo}\ \emph {et~al.}(2008)\citenamefont {Yogo},
  \citenamefont {Daido}, \citenamefont {Bulanov}, \citenamefont {Nemoto},
  \citenamefont {Oishi}, \citenamefont {Nayuki}, \citenamefont {Fujii},
  \citenamefont {Ogura}, \citenamefont {Orimo}, \citenamefont {Sagisaka},
  \citenamefont {Ma}, \citenamefont {Esirkepov}, \citenamefont {Mori},
  \citenamefont {Nishiuchi}, \citenamefont {Pirozhkov}, \citenamefont
  {Nakamura}, \citenamefont {Noda}, \citenamefont {Nagatomo}, \citenamefont
  {Kimura},\ and\ \citenamefont {Tajima}}]{yogo2008}%
  \BibitemOpen
  \bibfield  {author} {\bibinfo {author} {\bibfnamefont {A.}~\bibnamefont
  {Yogo}}, \bibinfo {author} {\bibfnamefont {H.}~\bibnamefont {Daido}},
  \bibinfo {author} {\bibfnamefont {S.~V.}\ \bibnamefont {Bulanov}}, \bibinfo
  {author} {\bibfnamefont {K.}~\bibnamefont {Nemoto}}, \bibinfo {author}
  {\bibfnamefont {Y.}~\bibnamefont {Oishi}}, \bibinfo {author} {\bibfnamefont
  {T.}~\bibnamefont {Nayuki}}, \bibinfo {author} {\bibfnamefont
  {T.}~\bibnamefont {Fujii}}, \bibinfo {author} {\bibfnamefont
  {K.}~\bibnamefont {Ogura}}, \bibinfo {author} {\bibfnamefont
  {S.}~\bibnamefont {Orimo}}, \bibinfo {author} {\bibfnamefont
  {A.}~\bibnamefont {Sagisaka}}, \bibinfo {author} {\bibfnamefont {J.~L.}\
  \bibnamefont {Ma}}, \bibinfo {author} {\bibfnamefont {T.~Z.}\ \bibnamefont
  {Esirkepov}}, \bibinfo {author} {\bibfnamefont {M.}~\bibnamefont {Mori}},
  \bibinfo {author} {\bibfnamefont {M.}~\bibnamefont {Nishiuchi}}, \bibinfo
  {author} {\bibfnamefont {A.~S.}\ \bibnamefont {Pirozhkov}}, \bibinfo {author}
  {\bibfnamefont {S.}~\bibnamefont {Nakamura}}, \bibinfo {author}
  {\bibfnamefont {A.}~\bibnamefont {Noda}}, \bibinfo {author} {\bibfnamefont
  {H.}~\bibnamefont {Nagatomo}}, \bibinfo {author} {\bibfnamefont
  {T.}~\bibnamefont {Kimura}}, \ and\ \bibinfo {author} {\bibfnamefont
  {T.}~\bibnamefont {Tajima}},\ }\href@noop {} {\bibfield  {journal} {\bibinfo
  {journal} {Physical Review E}\ }\textbf {\bibinfo {volume} {77}},\ \bibinfo
  {pages} {016401} (\bibinfo {year} {2008})}\BibitemShut {NoStop}%
\bibitem [{\citenamefont {Fukuda}\ \emph {et~al.}(2009)\citenamefont {Fukuda},
  \citenamefont {Faenov}, \citenamefont {Tampo}, \citenamefont {Pikuz},
  \citenamefont {Nakamura}, \citenamefont {Kando}, \citenamefont {Hayashi},
  \citenamefont {Yogo}, \citenamefont {Sakaki}, \citenamefont {Kameshima},
  \citenamefont {Pirozhkov}, \citenamefont {Ogura}, \citenamefont {Mori},
  \citenamefont {Esirkepov}, \citenamefont {Koga}, \citenamefont {Boldarev},
  \citenamefont {Gasilov}, \citenamefont {Magunov}, \citenamefont {Yamauchi},
  \citenamefont {Kodama}, \citenamefont {Bolton}, \citenamefont {Kato},
  \citenamefont {Tajima}, \citenamefont {Daido},\ and\ \citenamefont
  {Bulanov}}]{fukuda2009}%
  \BibitemOpen
  \bibfield  {author} {\bibinfo {author} {\bibfnamefont {Y.}~\bibnamefont
  {Fukuda}}, \bibinfo {author} {\bibfnamefont {A.~Y.}\ \bibnamefont {Faenov}},
  \bibinfo {author} {\bibfnamefont {M.}~\bibnamefont {Tampo}}, \bibinfo
  {author} {\bibfnamefont {T.~A.}\ \bibnamefont {Pikuz}}, \bibinfo {author}
  {\bibfnamefont {T.}~\bibnamefont {Nakamura}}, \bibinfo {author}
  {\bibfnamefont {M.}~\bibnamefont {Kando}}, \bibinfo {author} {\bibfnamefont
  {Y.}~\bibnamefont {Hayashi}}, \bibinfo {author} {\bibfnamefont
  {A.}~\bibnamefont {Yogo}}, \bibinfo {author} {\bibfnamefont {H.}~\bibnamefont
  {Sakaki}}, \bibinfo {author} {\bibfnamefont {T.}~\bibnamefont {Kameshima}},
  \bibinfo {author} {\bibfnamefont {A.~S.}\ \bibnamefont {Pirozhkov}}, \bibinfo
  {author} {\bibfnamefont {K.}~\bibnamefont {Ogura}}, \bibinfo {author}
  {\bibfnamefont {M.}~\bibnamefont {Mori}}, \bibinfo {author} {\bibfnamefont
  {T.~Z.}\ \bibnamefont {Esirkepov}}, \bibinfo {author} {\bibfnamefont
  {J.}~\bibnamefont {Koga}}, \bibinfo {author} {\bibfnamefont {A.~S.}\
  \bibnamefont {Boldarev}}, \bibinfo {author} {\bibfnamefont {V.~A.}\
  \bibnamefont {Gasilov}}, \bibinfo {author} {\bibfnamefont {A.~I.}\
  \bibnamefont {Magunov}}, \bibinfo {author} {\bibfnamefont {T.}~\bibnamefont
  {Yamauchi}}, \bibinfo {author} {\bibfnamefont {R.}~\bibnamefont {Kodama}},
  \bibinfo {author} {\bibfnamefont {P.~R.}\ \bibnamefont {Bolton}}, \bibinfo
  {author} {\bibfnamefont {Y.}~\bibnamefont {Kato}}, \bibinfo {author}
  {\bibfnamefont {T.}~\bibnamefont {Tajima}}, \bibinfo {author} {\bibfnamefont
  {H.}~\bibnamefont {Daido}}, \ and\ \bibinfo {author} {\bibfnamefont {S.~V.}\
  \bibnamefont {Bulanov}},\ }\href@noop {} {\bibfield  {journal} {\bibinfo
  {journal} {Physical Review Letters}\ }\textbf {\bibinfo {volume} {103}},\
  \bibinfo {pages} {165002} (\bibinfo {year} {2009})}\BibitemShut {NoStop}%
\bibitem [{\citenamefont {Bulanov}\ \emph {et~al.}(2010)\citenamefont
  {Bulanov}, \citenamefont {Bychenkov}, \citenamefont {Chvykov}, \citenamefont
  {Kalinchenko}, \citenamefont {Litzenberg}, \citenamefont {Matsuoka},
  \citenamefont {Thomas}, \citenamefont {Willingale}, \citenamefont {Yanovsky},
  \citenamefont {Krushelnick},\ and\ \citenamefont
  {Maksimchuk}}]{bulanov2010pop}%
  \BibitemOpen
  \bibfield  {author} {\bibinfo {author} {\bibfnamefont {S.~S.}\ \bibnamefont
  {Bulanov}}, \bibinfo {author} {\bibfnamefont {V.~Y.}\ \bibnamefont
  {Bychenkov}}, \bibinfo {author} {\bibfnamefont {V.}~\bibnamefont {Chvykov}},
  \bibinfo {author} {\bibfnamefont {G.}~\bibnamefont {Kalinchenko}}, \bibinfo
  {author} {\bibfnamefont {D.~W.}\ \bibnamefont {Litzenberg}}, \bibinfo
  {author} {\bibfnamefont {T.}~\bibnamefont {Matsuoka}}, \bibinfo {author}
  {\bibfnamefont {A.~G.~R.}\ \bibnamefont {Thomas}}, \bibinfo {author}
  {\bibfnamefont {L.}~\bibnamefont {Willingale}}, \bibinfo {author}
  {\bibfnamefont {V.}~\bibnamefont {Yanovsky}}, \bibinfo {author}
  {\bibfnamefont {K.}~\bibnamefont {Krushelnick}}, \ and\ \bibinfo {author}
  {\bibfnamefont {A.}~\bibnamefont {Maksimchuk}},\ }\href@noop {} {\bibfield
  {journal} {\bibinfo  {journal} {Physics of Plasmas}\ }\textbf {\bibinfo
  {volume} {17}},\ \bibinfo {pages} {043105} (\bibinfo {year}
  {2010})}\BibitemShut {NoStop}%
\bibitem [{\citenamefont {Sharma}(2018)}]{sharma2018high}%
  \BibitemOpen
  \bibfield  {author} {\bibinfo {author} {\bibfnamefont {A.}~\bibnamefont
  {Sharma}},\ }\href@noop {} {\bibfield  {journal} {\bibinfo  {journal}
  {Scientific reports}\ }\textbf {\bibinfo {volume} {8}},\ \bibinfo {pages}
  {2191} (\bibinfo {year} {2018})}\BibitemShut {NoStop}%
\bibitem [{\citenamefont {Fonseca}\ \emph {et~al.}(2002)\citenamefont
  {Fonseca}, \citenamefont {Silva}, \citenamefont {Tsung}, \citenamefont
  {Decyk}, \citenamefont {Lu}, \citenamefont {Ren}, \citenamefont {Mori},
  \citenamefont {Deng}, \citenamefont {Lee}, \citenamefont {Katsouleas},\ and\
  \citenamefont {Adam}}]{fonseca2002}%
  \BibitemOpen
  \bibfield  {author} {\bibinfo {author} {\bibfnamefont {R.~A.}\ \bibnamefont
  {Fonseca}}, \bibinfo {author} {\bibfnamefont {L.~O.}\ \bibnamefont {Silva}},
  \bibinfo {author} {\bibfnamefont {F.~S.}\ \bibnamefont {Tsung}}, \bibinfo
  {author} {\bibfnamefont {V.~K.}\ \bibnamefont {Decyk}}, \bibinfo {author}
  {\bibfnamefont {W.}~\bibnamefont {Lu}}, \bibinfo {author} {\bibfnamefont
  {C.}~\bibnamefont {Ren}}, \bibinfo {author} {\bibfnamefont {W.~B.}\
  \bibnamefont {Mori}}, \bibinfo {author} {\bibfnamefont {S.}~\bibnamefont
  {Deng}}, \bibinfo {author} {\bibfnamefont {S.}~\bibnamefont {Lee}}, \bibinfo
  {author} {\bibfnamefont {T.}~\bibnamefont {Katsouleas}}, \ and\ \bibinfo
  {author} {\bibfnamefont {J.~C.}\ \bibnamefont {Adam}},\ }\href@noop {}
  {\bibfield  {journal} {\bibinfo  {journal} {Lecture Notes in Computer
  Science}\ }\textbf {\bibinfo {volume} {2331}},\ \bibinfo {pages} {342}
  (\bibinfo {year} {2002})}\BibitemShut {NoStop}%
\bibitem [{\citenamefont {Katsouleas}\ \emph {et~al.}(1987)\citenamefont
  {Katsouleas} \emph {et~al.}}]{katsouleas1987}%
  \BibitemOpen
  \bibfield  {author} {\bibinfo {author} {\bibfnamefont {T.}~\bibnamefont
  {Katsouleas}} \emph {et~al.},\ }\href@noop {} {\bibfield  {journal} {\bibinfo
   {journal} {Particle Accelerators}\ }\textbf {\bibinfo {volume} {22}},\
  \bibinfo {pages} {81} (\bibinfo {year} {1987})}\BibitemShut {NoStop}%
\bibitem [{\citenamefont {Tzoufras}\ \emph {et~al.}(2008)\citenamefont
  {Tzoufras} \emph {et~al.}}]{tzoufras2008}%
  \BibitemOpen
  \bibfield  {author} {\bibinfo {author} {\bibfnamefont {M.}~\bibnamefont
  {Tzoufras}} \emph {et~al.},\ }\href@noop {} {\bibfield  {journal} {\bibinfo
  {journal} {Physical Review Letters}\ }\textbf {\bibinfo {volume} {101}},\
  \bibinfo {pages} {145002} (\bibinfo {year} {2008})}\BibitemShut {NoStop}%
\bibitem [{\citenamefont {Debayle}\ \emph {et~al.}(2017)\citenamefont
  {Debayle}, \citenamefont {Mollica}, \citenamefont {Vauzour}, \citenamefont
  {Wan}, \citenamefont {Flacco}, \citenamefont {Malka}, \citenamefont
  {Davoine},\ and\ \citenamefont {Gremillet}}]{debayle2017}%
  \BibitemOpen
  \bibfield  {author} {\bibinfo {author} {\bibfnamefont {A.}~\bibnamefont
  {Debayle}}, \bibinfo {author} {\bibfnamefont {F.}~\bibnamefont {Mollica}},
  \bibinfo {author} {\bibfnamefont {B.}~\bibnamefont {Vauzour}}, \bibinfo
  {author} {\bibfnamefont {Y.}~\bibnamefont {Wan}}, \bibinfo {author}
  {\bibfnamefont {A.}~\bibnamefont {Flacco}}, \bibinfo {author} {\bibfnamefont
  {V.}~\bibnamefont {Malka}}, \bibinfo {author} {\bibfnamefont
  {X.}~\bibnamefont {Davoine}}, \ and\ \bibinfo {author} {\bibfnamefont
  {L.}~\bibnamefont {Gremillet}},\ }\href@noop {} {\bibfield  {journal}
  {\bibinfo  {journal} {New Journal of Physics}\ }\textbf {\bibinfo {volume}
  {19}},\ \bibinfo {pages} {123013} (\bibinfo {year} {2017})}\BibitemShut
  {NoStop}%
\bibitem [{\citenamefont {Liu}\ \emph {et~al.}(2013)\citenamefont {Liu},
  \citenamefont {Wang}, \citenamefont {Liu}, \citenamefont {Fu}, \citenamefont
  {Xu}, \citenamefont {Yan},\ and\ \citenamefont {He}}]{liu2013}%
  \BibitemOpen
  \bibfield  {author} {\bibinfo {author} {\bibfnamefont {B.}~\bibnamefont
  {Liu}}, \bibinfo {author} {\bibfnamefont {H.~Y.}\ \bibnamefont {Wang}},
  \bibinfo {author} {\bibfnamefont {J.}~\bibnamefont {Liu}}, \bibinfo {author}
  {\bibfnamefont {L.~B.}\ \bibnamefont {Fu}}, \bibinfo {author} {\bibfnamefont
  {Y.~J.}\ \bibnamefont {Xu}}, \bibinfo {author} {\bibfnamefont {X.~Q.}\
  \bibnamefont {Yan}}, \ and\ \bibinfo {author} {\bibfnamefont {X.~T.}\
  \bibnamefont {He}},\ }\href@noop {} {\bibfield  {journal} {\bibinfo
  {journal} {Physical Review Letters}\ }\textbf {\bibinfo {volume} {110}},\
  \bibinfo {pages} {045002} (\bibinfo {year} {2013})}\BibitemShut {NoStop}%
\bibitem [{\citenamefont {Fiuza}\ \emph {et~al.}(2013)\citenamefont {Fiuza},
  \citenamefont {Stockem}, \citenamefont {Boella}, \citenamefont {Fonseca},
  \citenamefont {Silva}, \citenamefont {Haberberger}, \citenamefont
  {Tochitsky}, \citenamefont {Mori},\ and\ \citenamefont
  {Joshi}}]{fiuza2013pop}%
  \BibitemOpen
  \bibfield  {author} {\bibinfo {author} {\bibfnamefont {F.}~\bibnamefont
  {Fiuza}}, \bibinfo {author} {\bibfnamefont {A.}~\bibnamefont {Stockem}},
  \bibinfo {author} {\bibfnamefont {E.}~\bibnamefont {Boella}}, \bibinfo
  {author} {\bibfnamefont {R.~A.}\ \bibnamefont {Fonseca}}, \bibinfo {author}
  {\bibfnamefont {L.~O.}\ \bibnamefont {Silva}}, \bibinfo {author}
  {\bibfnamefont {D.}~\bibnamefont {Haberberger}}, \bibinfo {author}
  {\bibfnamefont {S.}~\bibnamefont {Tochitsky}}, \bibinfo {author}
  {\bibfnamefont {W.~B.}\ \bibnamefont {Mori}}, \ and\ \bibinfo {author}
  {\bibfnamefont {C.}~\bibnamefont {Joshi}},\ }\href@noop {} {\bibfield
  {journal} {\bibinfo  {journal} {Physics of Plasmas}\ }\textbf {\bibinfo
  {volume} {20}},\ \bibinfo {pages} {056304} (\bibinfo {year}
  {2013})}\BibitemShut {NoStop}%
\bibitem [{\citenamefont {Forslund}\ and\ \citenamefont
  {Shonk}(1970)}]{forslund1970csa}%
  \BibitemOpen
  \bibfield  {author} {\bibinfo {author} {\bibfnamefont {D.~W.}\ \bibnamefont
  {Forslund}}\ and\ \bibinfo {author} {\bibfnamefont {C.~R.}\ \bibnamefont
  {Shonk}},\ }\href@noop {} {\bibfield  {journal} {\bibinfo  {journal}
  {Physical Review Letters}\ }\textbf {\bibinfo {volume} {25}},\ \bibinfo
  {pages} {1699} (\bibinfo {year} {1970})}\BibitemShut {NoStop}%
\bibitem [{\citenamefont {Lu}\ \emph {et~al.}(2006)\citenamefont {Lu},
  \citenamefont {Huang}, \citenamefont {Zhou}, \citenamefont {Mori},\ and\
  \citenamefont {Katsouleas}}]{lu2006}%
  \BibitemOpen
  \bibfield  {author} {\bibinfo {author} {\bibfnamefont {W.}~\bibnamefont
  {Lu}}, \bibinfo {author} {\bibfnamefont {C.}~\bibnamefont {Huang}}, \bibinfo
  {author} {\bibfnamefont {M.}~\bibnamefont {Zhou}}, \bibinfo {author}
  {\bibfnamefont {W.~B.}\ \bibnamefont {Mori}}, \ and\ \bibinfo {author}
  {\bibfnamefont {T.}~\bibnamefont {Katsouleas}},\ }\href@noop {} {\bibfield
  {journal} {\bibinfo  {journal} {Physical Review Letters}\ }\textbf {\bibinfo
  {volume} {96}},\ \bibinfo {pages} {165002} (\bibinfo {year}
  {2006})}\BibitemShut {NoStop}%
\bibitem [{\citenamefont {Pukhov}\ and\ \citenamefont {Meyer-ter
  Vehn}(2002)}]{pukhov2002}%
  \BibitemOpen
  \bibfield  {author} {\bibinfo {author} {\bibfnamefont {A.}~\bibnamefont
  {Pukhov}}\ and\ \bibinfo {author} {\bibfnamefont {J.}~\bibnamefont {Meyer-ter
  Vehn}},\ }\href@noop {} {\bibfield  {journal} {\bibinfo  {journal} {Applied
  Physics B}\ }\textbf {\bibinfo {volume} {74}},\ \bibinfo {pages} {355}
  (\bibinfo {year} {2002})}\BibitemShut {NoStop}%
\bibitem [{\citenamefont {Wan}\ \emph {et~al.}(2018)\citenamefont {Wan},
  \citenamefont {Pai}, \citenamefont {Hua}, \citenamefont {Wu}, \citenamefont
  {Lu}, \citenamefont {Li}, \citenamefont {Zhang}, \citenamefont {Xu},
  \citenamefont {Joshi},\ and\ \citenamefont {Mori}}]{wan2018tri}%
  \BibitemOpen
  \bibfield  {author} {\bibinfo {author} {\bibfnamefont {Y.}~\bibnamefont
  {Wan}}, \bibinfo {author} {\bibfnamefont {C.-H.}\ \bibnamefont {Pai}},
  \bibinfo {author} {\bibfnamefont {J.}~\bibnamefont {Hua}}, \bibinfo {author}
  {\bibfnamefont {Y.}~\bibnamefont {Wu}}, \bibinfo {author} {\bibfnamefont
  {W.}~\bibnamefont {Lu}}, \bibinfo {author} {\bibfnamefont {F.}~\bibnamefont
  {Li}}, \bibinfo {author} {\bibfnamefont {C.}~\bibnamefont {Zhang}}, \bibinfo
  {author} {\bibfnamefont {X.}~\bibnamefont {Xu}}, \bibinfo {author}
  {\bibfnamefont {C.}~\bibnamefont {Joshi}}, \ and\ \bibinfo {author}
  {\bibfnamefont {W.}~\bibnamefont {Mori}},\ }\href@noop {} {\bibfield
  {journal} {\bibinfo  {journal} {Physics of Plasmas}\ }\textbf {\bibinfo
  {volume} {25}},\ \bibinfo {pages} {073105} (\bibinfo {year}
  {2018})}\BibitemShut {NoStop}%
\bibitem [{\citenamefont {Sylla}\ \emph {et~al.}(2012)\citenamefont {Sylla},
  \citenamefont {Veltcheva}, \citenamefont {Kahaly}, \citenamefont {Flacco},\
  and\ \citenamefont {Malka}}]{sylla2012}%
  \BibitemOpen
  \bibfield  {author} {\bibinfo {author} {\bibfnamefont {M.}~\bibnamefont
  {Sylla}}, \bibinfo {author} {\bibfnamefont {M.}~\bibnamefont {Veltcheva}},
  \bibinfo {author} {\bibfnamefont {S.}~\bibnamefont {Kahaly}}, \bibinfo
  {author} {\bibfnamefont {A.}~\bibnamefont {Flacco}}, \ and\ \bibinfo {author}
  {\bibfnamefont {V.}~\bibnamefont {Malka}},\ }\href@noop {} {\bibfield
  {journal} {\bibinfo  {journal} {Review of Scientific Instruments}\ }\textbf
  {\bibinfo {volume} {83}},\ \bibinfo {pages} {033507} (\bibinfo {year}
  {2012})}\BibitemShut {NoStop}%
\bibitem [{\citenamefont {Pai}\ \emph {et~al.}(2005)\citenamefont {Pai},
  \citenamefont {Huang}, \citenamefont {Kuo}, \citenamefont {Lin},
  \citenamefont {Wang}, \citenamefont {Chen}, \citenamefont {Lee},\ and\
  \citenamefont {Lin}}]{Pai2005}%
  \BibitemOpen
  \bibfield  {author} {\bibinfo {author} {\bibfnamefont {C.~H.}\ \bibnamefont
  {Pai}}, \bibinfo {author} {\bibfnamefont {S.~Y.}\ \bibnamefont {Huang}},
  \bibinfo {author} {\bibfnamefont {C.~C.}\ \bibnamefont {Kuo}}, \bibinfo
  {author} {\bibfnamefont {M.~W.}\ \bibnamefont {Lin}}, \bibinfo {author}
  {\bibfnamefont {J.}~\bibnamefont {Wang}}, \bibinfo {author} {\bibfnamefont
  {S.~Y.}\ \bibnamefont {Chen}}, \bibinfo {author} {\bibfnamefont {C.~H.}\
  \bibnamefont {Lee}}, \ and\ \bibinfo {author} {\bibfnamefont {J.~Y.}\
  \bibnamefont {Lin}},\ }\href@noop {} {\bibfield  {journal} {\bibinfo
  {journal} {Physics of Plasmas}\ }\textbf {\bibinfo {volume} {12}},\ \bibinfo
  {pages} {070707} (\bibinfo {year} {2005})}\BibitemShut {NoStop}%
\bibitem [{\citenamefont {Pogorelsky}\ and\ \citenamefont
  {Ben-Zvi}(2014)}]{pogorelsky2014}%
  \BibitemOpen
  \bibfield  {author} {\bibinfo {author} {\bibfnamefont {I.~V.}\ \bibnamefont
  {Pogorelsky}}\ and\ \bibinfo {author} {\bibfnamefont {I.}~\bibnamefont
  {Ben-Zvi}},\ }\href@noop {} {\bibfield  {journal} {\bibinfo  {journal}
  {Plasma Physics and Controlled Fusion}\ }\textbf {\bibinfo {volume} {56}},\
  \bibinfo {pages} {084017} (\bibinfo {year} {2014})}\BibitemShut {NoStop}%
\end{thebibliography}


\end{document}